\RequirePackage{ifpdf}
\ifpdf 
\documentclass[pdftex]{sigma}
\else
\documentclass{sigma}
\fi

\begin{document}

\newcommand{\C}{\mathbb C}
\newcommand{\R}{\mathbb R}
\newcommand{\Z}{\mathbb Z}
\newcommand{\N}{\mathbb N}
%
\renewcommand{\d}{\prime}
\newcommand{\dd}{{\prime \prime}}
\renewcommand{\Re}{{\rm Re}\,}
\renewcommand{\Im}{{\rm Im}\,}

\allowdisplaybreaks

\renewcommand{\thefootnote}{$\star$}

\renewcommand{\PaperNumber}{015}

\FirstPageHeading

\ShortArticleName{Anharmonic Oscillators with Inf\/initely Many Real Eigenvalues and $\mathcal{PT}$-Symmetry}

\ArticleName{Anharmonic Oscillators with Inf\/initely\\ Many Real Eigenvalues and $\boldsymbol{\mathcal{PT}}$-Symmetry\footnote{This paper is a
contribution to the Proceedings of the 5-th Microconference
``Analytic and Algebraic Me\-thods~V''. The full collection is
available at
\href{http://www.emis.de/journals/SIGMA/Prague2009.html}{http://www.emis.de/journals/SIGMA/Prague2009.html}}}

\Author{Kwang C. SHIN}

\AuthorNameForHeading{K.C. Shin}

\Address{Department of Mathematics, University of West Georgia, Carrollton, GA, 30118, USA}
\Email{\href{mailto:kshin@westga.edu}{kshin@westga.edu}}

\ArticleDates{Received October 11, 2009, in f\/inal form January 28, 2010;  Published online February 03, 2010}

\Abstract{We study the eigenvalue problem $-u^\dd+V(z)u=\lambda u$ in the complex plane with the boundary condition that
$u(z)$ decays to zero as $z$ tends to inf\/inity along the two rays $\arg z=-\frac{\pi}{2}\pm \frac{2\pi}{m+2}$,
where $V(z)=-(iz)^m-P(iz)$ for complex-valued polynomials~$P$ of degree at most $m-1\geq 2$. We provide an
asymptotic formula for eigenvalues and  a~necessary and suf\/f\/icient condition for the anharmonic oscillator to have
inf\/initely many real eigenvalues.}


\Classification{34L20; 34L40}

\renewcommand{\thefootnote}{\arabic{footnote}}
\setcounter{footnote}{0}

\section{Introduction}
\label{introduction} Theory of the so-called $\mathcal{PT}$-symmetric Hamiltonians has been a very active research
area in recent years \cite{Bender, Bender-1,CGM,Dorey,Znojil2,Ali1,Shin,Shin2,Znojil1}. While they are not
self-adjoint in general, many $\mathcal{PT}$-symmetric Hamiltonians have real spectra
\cite{Bender,CGM,Dorey,Shin}. Self-adjointness of Hamiltonians is not physical requirement, but guarantees reality
of spectrum and conservation of probability. So one searches for non-self-adjoint Hamiltonians that have real
spectra. In this paper, we study the Schr\"odinger eigenvalue problems in the complex plane with complex
polynomial potentials under decaying boundary conditions along two f\/ixed rays to inf\/inity. We will prove that the
Schr\"odinger eigenvalue problem has inf\/initely many real eigenvalues if and only if the eigenvalue problem is
$\mathcal{PT}$-symmetric or it is a translation of a $\mathcal{PT}$-symmetric problem.

For an integer $m\geq3$ we consider the Schr\"odinger eigenvalue problem
\begin{equation}\label{ptsym}
H u(z):=\left[-\frac{d^2}{dz^2}-(iz)^m-P(iz)\right]u(z)=\lambda u(z),\quad\text{for some $\lambda\in\C$},
\end{equation}
with the boundary condition that
\begin{equation}\label{bdcond}
\text{$u(z)\rightarrow 0$ exponentially, as $z\rightarrow \infty$ along the two rays} \  \arg
(z)=-\frac{\pi}{2}\pm \frac{2\pi}{m+2},
\end{equation}
where $P$ is a polynomial of degree at most $m-1$ of the form
\[
P(z)=a_1z^{m-1}+a_2z^{m-2}+\cdots+a_{m},\qquad a:=(a_1,a_2,\ldots, a_m)\in\C^m.
\]

The existence of inf\/initely many eigenvalues is known. Sibuya \cite{Sibuya} showed that the eigenvalues of $H$ are
zeros of an entire function of order $\rho:=\frac{1}{2}+\frac{1}{m}\in(0, 1)$ and hence, by the Hadamard
factorization theorem (see, e.g., \cite{Conway}), there are inf\/initely many eigenvalues.

The  Hamiltonian $H$ in \eqref{ptsym} with the potential $V(z)$ under the boundary condition \eqref{bdcond} is
called {\it $\mathcal{PT}$-symmetric} if $\overline{V(-\overline{z})}=V(z)$, $z\in\C$. Note that
$V(z)=-(iz)^m-P(iz)$ is a $\mathcal{PT}$-symmetric potential if and only if $a\in\R^m$.

In this paper,  we will derive the following asymptotic expansion of the eigenvalues $\lambda_n$, ordered in the
order of their magnitude.
\begin{theorem}\label{the_main}
There exists $N=N(m)\in\Z$ such that the eigenvalues $\left\{\lambda_n\right\}_{n=N}^{\infty}$ of $H$  satisfy
that
\begin{equation}\label{asymp_eqn1}
\sum_{j=0}^{m+1}c_j(
{a})\lambda_n^{\rho-\frac{j}{m}}+O\left(\lambda_n^{-\rho}\right)=n+\frac{1}{2}\quad \text{as $n\to+\infty$},
\end{equation}
where $c_0( {a})= \pi^{-1}\sin\left(\frac{\pi}{m}\right)B\left(\frac{1}{2},\,1+\frac{1}{m}\right)$ and $c_1(a)=0$
and  $c_j( {a})$, $2\leq j\leq m+1,$ are explicit non-zero polynomials in the coefficients $ {a}$ of the
polynomial potential, defined in~\eqref{c_def}.
\end{theorem}
As a consequence of \eqref{asymp_eqn1}, we obtain a necessary and suf\/f\/icient condition for the anharmonic
oscillator $H$ to have inf\/initely many real eigenvalues as follows.
\begin{theorem}\label{equiv-for}
The anharmonic oscillator $H$ with a potential $V(z)$ has infinitely many real eigenvalues if and only if  $H$
with the potential $V(z-z_0)$ is $\mathcal{PT}$-symmetric for some $z_0\in\C$.

Moreover, no $H$ has infinitely many real and infinitely many non-real eigenvalues.
\end{theorem}
If $H$ is $\mathcal{PT}$-symmetric, then $u(z)$ is an
eigenfunction associated with the eigenvalue $\lambda$ if and only
if $\overline{u(-\overline{z})}$ is an eigenfunction associated
with the eigenvalue $\overline{\lambda}$. Thus, the eigenvalues of
a~$\mathcal{PT}$-symmetric Hamiltonian $H$ either are real or come
in complex conjugates. Also,
magnitude of the large eigenvalues is strictly increasing~\cite{Shin2}. Thus,
if $H$ is $\mathcal{PT}$-symmetric, then
 eigenvalues are all real with at most f\/initely many exceptions.

 This paper is organized as follows. In Section~\ref{prop_sect},
 we will introduce results of  Hille~\cite{Hille},
 Sibuya~\cite{Sibuya}, and Shin~\cite{Shin2,Shin3} on properties of the solutions of~\eqref{ptsym}. We will present asymptotics of solutions of
\eqref{ptsym}, introduce the spectral determinant of $H$, and provide an asymptotic expansion for the spectral
determinant in the sector where all but f\/initely many eigenvalues lie. In Section \ref{asymp_eigen},  we will
prove Theorem~\ref{the_main} where we use the asymptotic expansion of the spectral determinant. In Section~\ref{last_sect}, we will prove Theorem~\ref{equiv-for} and provide other   spectral results; expres\-sing~$\lambda_n$ as an asymptotic formula in $n$ and obtaining an asymptotic formula for nearest neighbor spacing of
eigenvalues, monotonicity of their magnitude, and more.

\section{Properties of the solutions}
\label{prop_sect}

In this section, we will introduce results of Hille \cite{Hille}, Sibuya \cite{Sibuya}, and Shin
\cite{Shin2,Shin3} on the properties of solutions of \eqref{ptsym}.

First, let $u$ be a solution of (\ref{ptsym}) and let $v(z)=u(-iz)$. Then $v$ solves
\begin{equation}\label{rotated}
-v^\dd(z)+[z^m+P(z)+\lambda]v(z)=0.
\end{equation}
Solutions of \eqref{rotated} have simple asymptotic behavior near inf\/inity in the complex plane  and in order to
describe the asymptotic behavior, we will use the  Stokes sectors. The  Stokes sectors $S_k$ of the equation
(\ref{rotated}) are
 \[
 S_k=\left\{z\in \C:\left|\arg (z)-\frac{2k\pi}{m+2}\right|<\frac{\pi}{m+2}\right\}\quad\text{for}\  k\in \Z.
 \]
Hille \cite[\S~7.4]{Hille} showed that every nonconstant solution of (\ref{rotated}) either decays to zero or
blows up exponentially in each Stokes sector. Moreover, if a nonconstant solution $v$ of \eqref{rotated} decays in
a~Stokes sector, it must blow up in its two adjacent Stokes sectors. However, when $v$ blows up in a Stokes
sector, $v$ need not be decaying in its adjacent Stokes sectors. Thus, the decaying boundary condition
\eqref{bdcond} becomes that $v(z)\to 0$ as $z\to \infty$ in the sectors $S_1\cup S_{-1}$.

We will also use the following functions of the coef\/f\/icients $a\in\C^m$ of the polynomial $P$. Let
\begin{gather*}
\text{$b_{j,k}( {a})$ be the coef\/f\/icient of $z^{mk-j}$ in ${\frac{1}{2}\choose{k}}\left(P(z)\right)^k$\,\,for
$1\leq k\leq j$,\, and}
\\
b_j( {a})=\sum_{k=1}^j b_{j,k}( {a}),\quad j\in\N.
\end{gather*}
We also let $r_m=-\frac{m}{4}-\nu(a)$ and $\mu( {a})=\frac{m}{4}-\nu( {a})$, where
\begin{gather*}
\nu( {a})=\left\{
              \begin{array}{ll}
              0   &\text{if $m$ is odd,}\\
              b_{\frac{m}{2}+1}( {a})   &\text{if $m$ is even.}
              \end{array}
                         \right.
\end{gather*}

Now we are ready to introduce results of Sibuya \cite{Sibuya} and Shin \cite{Shin3} on existence of a solution
that has a f\/ixed asymptotic behavior in $S_{-1}\cup S_0\cup S_1$, and on its properties.
\begin{theorem}\label{prop}
Equation \eqref{rotated}, with $a\in \C^m$, has a solution  $f(z,a,\lambda)$ with the following properties.
\begin{enumerate}\itemsep=0pt
\item[$(i)$] $f(z,a,\lambda)$ is an entire function of $z$, $a $ and $\lambda$.
\item[$(ii)$] $f(z,a,\lambda)$ and $f^\d(z,a,\lambda)=\frac{\partial}{\partial z}f(z,a,\lambda)$ admit the following asymptotic
expansions. For each fixed $\varepsilon>0$,
\begin{gather*}
f(z,a,\lambda)=   z^{r_m}\big(1+O\big(z^{-1/2}\big)\big)\exp\left[-F(z,a,\lambda)\right],\\
f^\d(z,a,\lambda)= -z^{r_m+\frac{m}{2}}\big(1+O\big(z^{-1/2}\big)\big)\exp\left[-F(z,a,\lambda) \right],
\end{gather*}
as $z$ tends to infinity in  the sector $|\arg z|\leq \frac{3\pi}{m+2}-\varepsilon$, uniformly on each compact set
of $(a,\lambda)$-values. Here
\[
F(z,a,\lambda)=\frac{2}{m+2}z^{\frac{m}{2}+1}+\sum_{1\leq j<\frac{m}{2}+1}\frac{2}{m+2-2j}b_j(a) z^{\frac{1}{2}(m+2-2j)}.
\]
\item[$(iii)$] For each  $\delta>0$, $f$ and $f^\d$ also admit the asymptotic expansions,
\begin{gather}
f(0,a,\lambda)= \left[1+O\left(\lambda^{-\rho}\right)\right]\lambda^{-1/4}\exp\left[L(a,\lambda)\right],\label{eq1}\\
f^\d(0,a,\lambda)= -\left[1+O\left(\lambda^{-\rho}\right)\right]\lambda^{1/4}\exp\left[L(a,\lambda)\right],\label{eq2}
\end{gather}
as $\lambda\to\infty$ in the sector $|\arg(\lambda)|\leq\pi-\delta$, uniformly on each compact set of $a\in\C^m$,
where $\rho:=\frac{1}{2}+\frac{1}{m}$
\[
L(a,\lambda)=\int_0^{+\infty}\left(\sqrt{t^m+P(t)+\lambda}-
t^{\frac{m}{2}}-\sum_{j=1}^{\lfloor\frac{m+1}{2}\rfloor}b_j(a)t^{\frac{m}{2}-j}
                   -\frac{\nu(a)}{t+1}\right)dt.
\]
\item[$(iv)$] The entire functions  $\lambda\mapsto f(0,a,\lambda)$
and $\lambda\mapsto f^\d(0,a,\lambda)$ have orders
$\rho=\frac{1}{2}+\frac{1}{m}$.
\end{enumerate}
\end{theorem}

\begin{proof}
See Theorems 6.1, 7.2, 19.1, and 20.1 in \cite{Sibuya} for proof. Sibuya showed \eqref{eq1} and \eqref{eq2} with
error terms $o(1)$ and I improved them to $O\left(\lambda^{-\rho}\right)$ in \cite{Shin3}.
\end{proof}

\subsection{Spectral determinant}
In this subsection, we will introduce the function whose zeros are the eigenvalues of $H$.

Let
\[
\omega=\exp\left[\frac{2\pi i}{m+2}\right]
\]
and def\/ine
\[
G^k(a):=\big(\omega^{-k}a_1, \omega^{-2k}a_2,\ldots,\omega^{-mk}a_m\big)\quad \text{for} \  k\in \Z.
\]
Then one can see from the def\/inition of $b_{j,k}( {a})$ that
\begin{equation}\label{bjk}
b_{j,k}\big(G^{\ell}( {a})\big)=\omega^{-j\ell}b_{j,k}( {a})\qquad \text{and}\qquad  b_j\big(G^{\ell}( {a})\big)=\omega^{-j\ell}b_j(
{a})\quad \text{for} \  \ell\in\Z.
\end{equation}

Also, by Theorem \ref{prop} there exists $f(z,G^k(a),\omega^{-mk}\lambda)$ that decays in $S_0$ and we let
\[
f_k(z,a,\lambda):=f\big(\omega^{-k}z,G^k(a),\omega^{-mk}\lambda\big).
\]
So $f_k$ decays in $S_k$ and blows up in its two adjacent Stokes sectors. Since $f_0$ decays in $S_0$ while~$f_1$
blows up in the same sector, we conclude that $f_0$ and $f_1$ are linearly independent. Thus, any solution can be
expressed as a linear combination of these two. In particular,
\[
f_{-1}(z,a,\lambda)=C(a,\lambda)f_0(z,a,\lambda)+\widetilde{C}(a,\lambda)f_{1}(z,a,\lambda)\quad \text{for some
$C(a,\lambda)$ and $\widetilde{C}(a,\lambda)$.}
\]
One can write these coef\/f\/icients in terms of the Wronskians of solutions. That is,
\[
C(a,\lambda)=\frac{\mathcal{W}_{-1,1}(a,\lambda)}{\mathcal{W}_{0,1}(a,\lambda)}\qquad\text{and}\qquad
\widetilde{C}(a,\lambda)= -\frac{\mathcal{W}_{-1,0}(a,\lambda)}{\mathcal{W}_{0,1}(a,\lambda)},
\]
where $\mathcal{W}_{j,k}=f_jf_k^\d -f_j^\d f_k$ is the Wronskian of $f_j$ and $f_k$. Since $f_0$ and $f_1$ are
linearly independent, $\mathcal{W}_{0,1}(a,\lambda)\not=0$ and likewise $\mathcal{W}_{-1,0}(a,\lambda)\not=0$.
Then by Theorem \ref{prop}$(i)$, for each f\/ixed $a\in\C^m$, the function $\lambda\mapsto C(a,\lambda)$ is  entire.
Moreover, $\lambda$ is an eigenvalue of $H$ if and only if $C(a,\lambda)=0$. We call $C(a,\lambda)$ the spectral
determinant of $H$.

\subsection[Asymptotic expansions of $C(a,\lambda)$]{Asymptotic expansions of $\boldsymbol{C(a,\lambda)}$}\label{asymp_sect}

In \cite{Shin2}, the asymptotics of $C(a,\lambda)$ is provided and it is showed that all but f\/initely many zeros
$\lambda$ of $C(a,\lambda)$ lie in a small sector containing the positive real $\lambda$-axis. Here we will
improve this asymptotics of $C(a,\lambda)$ as $\lambda\to\infty$ in a sector containing the positive real
$\lambda$-axis as follows.
\begin{theorem}\label{thm_sector2}
Suppose that $m\geq 4$. Then for each f\/ixed $a\in\C^m$ and $0<\delta<\frac{\pi}{m+2},$
\begin{gather}
C(a,\lambda)= \big[\omega^{\frac{1}{2}}+O\left(\lambda^{-\rho}\right)\big]\exp\left[L\big(G^{-1}(a),\omega^{-2}\lambda\big)-L(a,\lambda)\right]\nonumber\\
\phantom{C(a,\lambda)=}{} +\big[\omega^{\frac{1}{2}+2\nu(a)}+O\left(\lambda^{-\rho}\right)\big]\exp\left[L(G(a),\omega^{2}\lambda)-L(a,\lambda)\right],\label{asymp_1}
\end{gather}
as $\lambda\to\infty$ in the sector
\begin{equation}\label{sector4}
\pi-\frac{4\lfloor\frac{m}{2}\rfloor\pi}{m+2}+\delta \leq \arg(\lambda)\leq \pi-\frac{4\pi}{m+2}-\delta.
\end{equation}
If $m=3$ then
\begin{gather}
C(a,\lambda)= \big[-\omega^{-2}+O\left(\lambda^{-\rho}\right)\big]\exp\left[L\big(G^{4}(a),\omega^{-2}\lambda\big)-L(a,\lambda)\right]\nonumber\\
\phantom{C(a,\lambda)=}{} -\big[i\omega^{\frac{7}{4}}+O\left(\lambda^{-\rho}\right)\big]\exp\left[-L(G^2(a),\omega^{-1}\lambda)-L(a,\lambda)\right],\label{asymp_2}
\end{gather}
as $\lambda\to\infty$ in the sector
\begin{equation}\label{sector2-1}
-\frac{\pi}{5}+\delta \leq \arg(\lambda)\leq \pi-\delta.
\end{equation}
\end{theorem}

\begin{proof}
In Theorems 13 and 14 of Shin \cite{Shin2},  similar asymptotics
of $C(a,\lambda)$ are proved with the error terms $o(1)$ instead
of $O\left(\lambda^{-\rho}\right)$. With the improved asymptotics
\eqref{eq1} and \eqref{eq2}, one can closely follow proofs of
Theorems 13 and 14 in \cite{Shin2} to complete the proof.
\end{proof}

Next, in order to further examine the improved asymptotics of $C(a,\lambda)$ in Theorem \ref{thm_sector2}, we will
use the following asymptotics of $L(a,\lambda)$.

\begin{lemma}\label{asy_lemma}
Let $m\geq 3$ and $a\in\C^m$ be fixed. Then
\[
L(a,\lambda)=\sum_{j=0}^{\infty}K_{m,j}(a)\lambda^{\frac{1}{2}+\frac{1-j}{m}}-\frac{\nu(a)}{m}\ln(\lambda)
\]
as $\lambda\to\infty$ in the sector $|\arg(\lambda)|\leq \pi-\delta$ with
$K_{m,0}(a)=\frac{B\left(\frac{1}{2},1+\frac{1}{m}\right)}{2\cos\left(\frac{\pi}{m}\right)}>0$ and for $j\geq 1$,
\begin{equation}\label{Kmj=}
K_{m,j}( {a})=\sum_{k=\lfloor\frac{j-1}{m}\rfloor+1}^{j}K_{m,j,k}\,b_{j,k}( {a}),
\end{equation}
where for $\frac{j}{m}\leq k\leq j$,
\begin{gather*}
K_{m,j,k} :=\left\{
                    \begin{array}{ll}
\displaystyle -\frac{2}{m}
\quad &\text{if $j=1$},\vspace{2mm}\\
\displaystyle \frac{2}{m}\left(\ln 2-\sum\limits_{s=1}^{k-1}\frac{1}{2s-1}\right) \, &\displaystyle \text{if $j=\frac{m}{2}+1$, $m$ even,}\vspace{2mm}\\
\displaystyle                  \frac{1}{m}B\left(k-\frac{j-1}{m},\,\frac{j-1}{m}-\frac{1}{2}\right)  &\displaystyle
                  \text{if  $j\not=1$ or $j\not=\frac{m}{2}+1$.}
\end{array}\right.
\end{gather*}
\end{lemma}
\begin{proof}
See the Appendix in  \cite{Shin3} for a proof.
\end{proof}
\begin{remark}\label{remark1}\sloppy
The coef\/f\/icients $K_{m,j}( {a})$ have the following properties that can be deduced from~\eqref{Kmj=} and the
def\/inition of $b_{j,k}(a)$:
\begin{itemize}\itemsep=0pt
\item[$(i)$] The $K_{m,j}( {a})$ are all real polynomials in terms
of the coef\/f\/icients $ {a}$ of $P$.
\item[$(ii)$] For each
$1\leq j\leq m$, the polynomial $K_{m,j}( {a})$ depends only on
$a_1,a_2,\dots,a_{j}$. Furthermore, it is a non-constant linear
function of $a_j$.
\end{itemize}
\end{remark}

\section{Proof of Theorem \ref{the_main}}\label{asymp_eigen}

In this section, we will prove Theorem \ref{the_main}.
\begin{proof}[Proof of Theorem~\ref{the_main}]
In order to get asymptotics for the eigenvalues $\lambda$, we
examine asymptotics of  $C(a,\lambda)$ since the zeros of
$C(a,\lambda)$ are the eigenvalues of $H$. Since all but f\/initely
many eigenvalues lie near the positive real $\lambda$-axis, the
asymptotics in Theorem \ref{thm_sector2} would be suf\/f\/icient for
our purpose.

For $m\geq 4$ and $a\in\C^m$ f\/ixed, we set $C(a,\lambda)=0$ in
\eqref{asymp_1}, rearrange the asymptotic equation to get
\[
\big[1+O\left(\lambda^{-\rho}\right)\big]\exp\left[L\big(G(a),\omega^2\lambda\big)-L\big(G^{-1}(a),\omega^{-2}\lambda\big)\right]
=-\omega^{-2\nu(a)},
\]
and absorb the term $[1+O\left(\lambda^{-\rho}\right)]$
 into the exponential term to obtain
\[
\exp\left[L\big(G(a),\omega^2\lambda\big)-L\big(G^{-1}(a),\omega^{-2}\lambda\big)+O\left(\lambda^{-\rho}\right)\right]
=-\omega^{-2\nu(a)},
\]
or equivalently
\begin{equation}\label{main_eq}
L\big(G(a),\omega^2\lambda\big)-L\big(G^{-1}(a),\omega^{-2}\lambda\big)+O\left(\lambda^{-\rho}\right) =(2n+1)\pi i
-\frac{4\nu(a)\pi}{m+2}i,\quad \text{for some $n\in\Z$}.
\end{equation}

 Since $L(a,\lambda)=K_{m,0}(a)\lambda^{\rho}(1+o(1))$ by Lemma \ref{asy_lemma}, one gets
\begin{equation}\label{L-estmat}
L\big(G(a),\omega^2\lambda\big)-L\big(G^{-1}(a),\omega^{-2}\lambda\big)+O\left(\lambda^{-\rho}\right)
=2iK_{m,0}(a)\sin\left(\frac{2\pi}{m}\right)\lambda^{\rho}(1+o(1)),
\end{equation}
as $\lambda\to\infty$ in the sector \eqref{sector4} containing the
positive real $\lambda$-axis. Thus, the function
\[
\lambda\mapsto H(a,\lambda):=L(G(a),\omega^2\lambda)
-L(G^{-1}(a),\omega^{-2}\lambda)+O\left(\lambda^{-\rho}\right)
\]
 maps the region $|\lambda|\geq M$ and $|\arg(\lambda)|\leq
\varepsilon$ for some large $M>0$ and small $\varepsilon>0$ onto a region that contains the positive imaginary
axis near inf\/inity. Also, $H(a,\cdot)$ is analytic in the region $|\lambda|\geq M$ and $|\arg(\lambda)|\leq
\varepsilon$. Then, Sibuya \cite[pp.~131--133]{Sibuya} showed that for every closed ball $\{a\in\C^m: |a|\leq
r\}$, there exists $N_0=N_0(m, r)\in\N$ such that
 for every integer $n\geq N_0$ there exists exactly one $\lambda_n$ for which \eqref{main_eq} holds, that is,
\begin{equation}\label{L-eq}
H(a,\lambda_n)=(2n+1)\pi i -\frac{4\nu(a)\pi}{m+2}i.
\end{equation}
(Sibuya showed the existence of such an $N_0$ with the estimate \eqref{L-estmat}.)

Hence, from Lemma \ref{asy_lemma} we infer that
\begin{gather}
H(a,\lambda_n)= L\big(G(a),\omega^2\lambda_n\big)-L\big(G^{-1}(a),\omega^{-2}\lambda_n\big)+O\left(\lambda_n^{-\rho}\right)\nonumber\\
\phantom{H(a,\lambda_n)}{} =  \sum_{j=0}^{m+1}\left[ K_{m,j}(G(a))\left(\omega^{2}\lambda_n\right)^{\frac{1}{2}+\frac{1-j}{m}}
-K_{m,j}(G^{-1}(a))(\omega^{-2}\lambda_n)^{\frac{1}{2}+\frac{1-j}{m}}\right]\nonumber\\
\phantom{H(a,\lambda_n)=}{} +\frac{\nu\left(G^{-1}(a)\right)}{m}\ln\left(\omega^{-2}\lambda_n\right)-\frac{\nu\left(G(a)\right)}{m}\ln\left(\omega^{2}\lambda_n\right)
+O\left(\lambda_n^{-\rho}\right)\nonumber\\
\phantom{H(a,\lambda_n)}{}= 2i\sum_{j=0}^{m+1}K_{m,j}(a)\sin\left(\frac{2(1-j)\pi}{m}\right)\lambda_n^{\frac{1}{2}+\frac{1-j}{m}}+\frac{\nu(a)}{m}\frac{8\pi
i}{m+2}+O\left(\lambda_n^{-\rho}\right),\label{K-eq}
\end{gather}
where we used $\nu\left(G^{\pm1}(a)\right)=-\nu(a)$ and by  \eqref{bjk} and \eqref{Kmj=}, for $1\leq j\leq m+1$,
\begin{gather}
 K_{m,j}(G(a))\big(\omega^{2}\big)^{\frac{1}{2}+\frac{1-j}{m}}-K_{m,j}\big(G^{-1}(a)\big)\big(\omega^{-2}\big)^{\frac{1}{2}+\frac{1-j}{m}}\nonumber\\
\qquad{}=K_{m,j}(a)\omega^{-j}\big(\omega^{2}\big)^{\frac{1}{2}+\frac{1-j}{m}}-K_{m,j}(a)\omega^j\big(\omega^{-2}\big)^{\frac{1}{2}+\frac{1-j}{m}}\nonumber\\
\qquad{}=K_{m,j}(a)\big(e^{(1-j)\frac{2\pi i}{m}}-e^{-(1-j)\frac{2\pi i}{m}}\big)\nonumber\\
\qquad{}=2iK_{m,j}(a)\sin\left(\frac{2(1-j)\pi}{m}\right).\label{K_eq}
\end{gather}
Next, one can combine \eqref{L-eq} and \eqref{K-eq} and rearrange the resulting equation to obtain
\eqref{asymp_eqn1}. That~is,
\[
\sum_{j=0}^{m+1}c_j(a)\lambda_n^{\frac{1}{2}+\frac{1-j}{m}}
+O\left(\lambda_n^{-\rho}\right)=n+\frac{1}{2}\quad \text{as
$n\to\infty$},
\]
where for $0\leq j\leq m+1$,
\begin{equation}\label{c_def}
c_j(a):=\left\{
                    \begin{array}{ll}
\displaystyle  \frac{1}{\pi}K_{m,j}(a)\sin\left(\frac{2(1-j)\pi}{m}\right)
  &\displaystyle\text{if $j\not=\frac{m}{2}+1$},\\
&\\
\displaystyle \frac{2\nu(a)}{m}   &\displaystyle \text{if $j=\frac{m}{2}+1$, $m$ even.}
\end{array}\right.
\end{equation}

We still need to examine how many eigenvalues of $H$ are not in the set $\{\lambda_n\}_{n\ge N_0}$ to complete our
proof. To do this we will use Hurwitz's theorem in complex analysis.

Hurwitz's theorem \cite{Conway} says that if a sequence of analytic functions $\varphi_k$ converges uniformly to an analytic
function $\varphi$ on any compact subsets of their common domain, then for all large~$k$ functions~$\varphi_k$ and
$\varphi$ have the same number of zeros in any open subset whose boundary does not contain any zeros of~$\varphi$.
 Since the eigenvalues are the zeros of $C(a,\lambda)$ that is an analytic function in variables $a$ and $\lambda$,
 Hurwitz's theorem implies that  eigenvalues varies continuously as~$a$ varies. That is, $\lambda_n(a)$ varies continuously from
$\lambda_n(\widehat{a})$ to $\lambda_n(\widetilde{a})$ as $a$ varies from $\widehat{a}$ to $\widetilde{a}$ in the
closed ball, where we used $\lambda_n(a)$  in order to clearly indicate its dependence on $a$.

 When $a=0$,  there exists an integer $N=N(m)\leq N_0$ such that the eigenvalues of $H$ that are not
 in $\{\lambda_n(0)\}_{n\geq N_0}$ is exactly $(N_0-N)$.
  Also, Hurwitz's theorem implies that when eigenvalues collide as $a$ varies, the number of eigenvalues before and after the
collision remains the same. So there is no sudden appearance or disappearance of the eigenvalues as $a$ varies and
hence, there are exactly $(N_0-N)$-eigenvalues that are not in the set $\{\lambda_n(a)\}_{n\geq N_0}$. Therefore,
the eigenvalues $\{\lambda_n(a)\}_{n\geq N}$ satisfy \eqref{asymp_eqn1}. This completes the proof for $m\geq 4$.

For $m=3$, proof is very similar to the case $m\geq 4$ above and we will use \eqref{asymp_2} in the place of~\eqref{asymp_1}. If $C(a,\lambda)=0$ then we rearrange the asymptotic formula \eqref{asymp_2} with
$C(a,\lambda)=0$ to obtain
\[
\big[1+O\left(\lambda^{-\rho}\right)\big]\exp\left[-L\big(G^{4}(a),\omega^{-2}\lambda\big)-L\big(G^{2}(a),\omega^{-1}\lambda\big)\right]=e^{\pi
i},
\]
as $\lambda\to\infty$ in the sector~\eqref{sector2-1}. Next, like
in \eqref{K_eq} we examine
\begin{gather*}
 -K_{3,j}\big(G^4(a)\big)\left(\omega^{-2}\lambda\right)^{\frac{1}{2}+\frac{1-j}{3}}-K_{3,j}\big(G^2(a)\big)
\left(\omega^{-1}\lambda\right)^{\frac{1}{2}+\frac{1-j}{3}}\nonumber\\
\qquad{}=-\big[\omega^{-4j}\omega^{-2\left(\frac{1}{2}+\frac{1-j}{3}\right)}+\omega^{-2j}\omega^{-\left(\frac{1}{2}+\frac{1-j}{3}\right)}\big]K_{3,j}(a)\lambda^{\frac{1}{2}+\frac{1-j}{3}}\nonumber\\
\qquad{} =-\big[e^{-\frac{4(j-1)}{3}\pi i}-e^{-\frac{2(j-1)}{3}\pi i}\big]K_{3,j}(a)\lambda^{\frac{1}{2}+\frac{1-j}{3}}\nonumber\\
\qquad{} =2i\sin\left(\frac{2(1-j)\pi}{3}\right)K_{3,j}(a)\lambda^{\frac{1}{2}+\frac{1-j}{3}},\nonumber
\end{gather*}
where we used \eqref{bjk}, \eqref{Kmj=}, and $\omega^{\frac{5}{2}}=-1$. Then one can complete proof for $m=3$ just like in the case of $m\geq 4$.
\end{proof}
\begin{remark}
{\rm We order the eigenvalues in the order of their magnitude and we showed that eigenvalues vary continuously as
 the coef\/f\/icients $a=(a_1,\dots, a_m)$ vary. For the large eigenvalues,  $\lambda_n(a)$ are continuous functions of $a\in\C^m$ (c.f. \eqref{mono}). However,
for small eigenvalues, even though eigenvalues varies continuously as $a$ varies, we do not claim that
$\lambda_n(a)$ ordered in their order of magnitude are continuous. While these small eigenvalues are continuously
moving in the complex plane, their indices can be switched and how we number these small eigenvalues does not
af\/fect the asymptotic result in Theorem~\ref{the_main}.}
\end{remark}

\section{Corollaries and proof of Theorem \ref{equiv-for}}\label{last_sect}
In this section, we will introduce some results deduced from Theorem \ref{the_main}.

First, we invert \eqref{asymp_eqn1}, expressing $\lambda_n$ as an asymptotic formula in $n$. Also, we obtain an
asymptotic formula for nearest neighbor spacing of eigenvalues and monotonicity of their magnitude.
\begin{corollary}
 One can compute numbers $d_j( {a})$ explicitly such that
\begin{equation}\label{somm_eq}
\lambda_n=\sum_{j=0}^{m+1}d_j( {a})\cdot
\left(n+\frac{1}{2}\right)^{\frac{2m}{m+2}\left(1-\frac{j}{m}\right)}+O\big(n^{-\frac{4}{m+2}}\big)\quad \text{as
$n\to+\infty$}.
\end{equation}
Also, the space between successive eigenvalues is
\[
 \lambda_{n+1}-\lambda_n
\underset{n\to+\infty}{=}\frac{2m}{m+2}\left(\frac{\pi}{B\left(\frac{1}{2},1+\frac{1}{m}\right)}\right)^{\frac{2m}{m+2}}\cdot
\left(n+\frac{1}{2}\right)^{\frac{m-2}{m+2}}+o\big(n^{\frac{m-2}{m+2}}\big)\quad \text{as $n\to+\infty$.}
\]

In particular, $\lim\limits_{n\to\infty}\left|\lambda_{n+1}-\lambda_n\right|=\infty$ and
$\lim\limits_{n\to+\infty}\arg(\lambda_n)=0.$ Hence:
\begin{equation}\label{mono}
\left|\lambda_n\right|<\left|\lambda_{n+1}\right|\quad \text{for all large $n$}.
\end{equation}
\end{corollary}

\begin{proof}
 Since \eqref{asymp_eqn1} is an asymptotic equation,  one can solve it for $\lambda_n$ to get \eqref{somm_eq}, where
$d_0(a)=\left(c_0\right)^{-\frac{2m}{m+2}}$ and $d_j( {a})$ for $1\leq j\leq m+1$ can be computed recursively and
explicitly. Then since $d_0>0$, one can deduce
\begin{equation}\label{eq_111}
\lim_{n\to0}\arg(\lambda_n)=0\qquad \text{and}\qquad \lim_{n\to\infty}|\lambda_n|=\infty.
\end{equation}
Also, we obtain
\begin{equation}\label{mono_11}
 \lambda_{n+1}
\underset{n\to+\infty}{=}\lambda_n+\frac{2m}{m+2}\left(d_0(a)\right)^{\frac{2m}{m+2}}\cdot
\left(n+\frac{1}{2}\right)^{\frac{m-2}{m+2}}+o\big(n^{\frac{m-2}{m+2}}\big)\quad \text{as
$n\to+\infty$,}
\end{equation}
where we use the generalized binomial expansion to $\big(1+\big(n+\frac{1}{2}\big)^{-1}\big)^{\alpha}$ for
some $\alpha\in\R$. Since $d_0>0$, equations~\eqref{eq_111} and~\eqref{mono_11} imply
$\lim_{n\to0}\left|\lambda_{n+1}-\lambda_n\right|=\infty$ and hence, \eqref{mono} holds.
\end{proof}

When $H$ is $\mathcal{PT}$-symmetric, the eigenvalues either are real or else appear in complex conjugates. Thus,
if $H$ is $\mathcal{PT}$-symmetric, then by monotonicity of magnitude of the large eigen\-values~\eqref{mono}, $H$
cannot have an inf\/inite pairs of complex conjugate eigenvalues and hence all but f\/initely many eigenvalues are
real.

Next, we will prove Theorem \ref{equiv-for} on a necessary and suf\/f\/icient condition for an anharmonic oscillator
$H$ having inf\/initely many real eigenvalues.
\begin{proof}[Proof of Theorem~\ref{equiv-for}]
Suppose that $H$ or its translation is $\mathcal{PT}$-symmetric. It is easy to check that  the potentials $V(z)$
and $V(z-z_0)$ generate the same set of eigenvalues for any $z_0\in\C$. Thus, since $H$ or its translation is
$\mathcal{PT}$-symmetric, eigenvalues of $H$ either are real or come in complex conjugates. Then by~\eqref{mono},
we conclude that all but f\/initely many eigenvalues are real. Since there are inf\/initely many eigenvalues
\cite{Sibuya} as we noted in the Introduction, there are inf\/initely many real eigenvalues.

{\sloppy
Suppose that $H$ has inf\/initely many real eigenvalues. Then
f\/irst, from the properties of~$K_{m,j}(a)$ in 
Remark~\ref{remark1} 
and \eqref{c_def}, one can derive
the following properties of $c_j(a)$:
\begin{itemize}\itemsep=0pt
\item[$(i)$] The $c_j( {a})$ are all real-valued polynomials in
terms of the coef\/f\/icients $ {a}$ of $P$.
\item[$(ii)$] For
$2\leq j\leq m$, the polynomial $c_j( {a})$ depends only on $a_1,a_2,\dots,a_{j}$. Furthermore, it is a~non-constant linear function of $a_j$.
\end{itemize}

}

Notice that if $c_j(a)\not\in\R$ for some $0\leq j\leq m+1$, then \eqref{asymp_eqn1} implies that all but f\/initely
many eigenvalues are non-real. Thus, if $H$ has inf\/initely many real eigenvalues, then $c_j(a)\in\R$ for all
$0\leq j\leq m+1$.

Recall that $c_0(a)$ is real and depends only on $m$, and $c_1(a)=0$. Next, if $a_1\not=0$, then we can choose
$z_0=-\frac{a_1}{m}i$ and  we consider the translation of $H$, replacing the potential $V(z)$ by $V(z-z_0)$. The
translated $H$ has its potential with a~vanishing $z^{m-1}$-term (i.e., $a_1=0$). Then by the above properties $(i)$
and $(ii)$ of $c_2(a), c_2(a)\in\R$ is a real polynomial in $a_1$ and $a_2$, and it is a nonconstant linear
function in $a_2$. Thus, since $a_1=0\in\R$ and $c_2(a)\in\R$, we obtain $a_2\in\R$. Likewise, since $a_1, a_2 ,
c_3(a)\in\R$, by the above properties of $c_3(a), a_3\in\R$ and hence, recursively we conclude that $a\in\R^m$.
Therefore, the translated $H$ is $\mathcal{PT}$-symmetric.
\end{proof}

\pdfbookmark[1]{References}{ref}
\LastPageEnding

\end{document}